# Analytical Model for Calculating Gain Pattern of Antennas Implanted in Large Host Bodies

Mingxiang Gao, Zvonimir Šipuš, Icaro V. Soares, Sujith Raman, Denys Nikolayev, and Anja K. Skrivervik

*Abstract*—This paper presents a method for the fast and accurate estimation of the gain pattern and maximum gain of an implanted antenna including the effect of the host body, under the assumption that the latter is electrically large. The estimation procedure is based on the radiation of an elementary dipole source placed in a planar body model. The derivation of closed-form expressions is based on spherical wave analysis and the Green's functions for layered media. The validity of this approximation for practical cases is shown on different implanted antennas, where the results are compared to full wave simulations and measurements.

## I. INTRODUCTION

Implantable bioelectronics and biosensors have enabled important advances in biomedical research and applications, including but not limited to cardiac pacing, physiological monitoring, drug delivery, and neurostimulation [1], [2]. To enable wireless communication and operation, wireless techniques are used in various implantable devices to transmit and receive data from external equipment. However, the presence of the host body significantly affects the wireless link efficiency, due to the lossy behavior of most biological tissues respect to electromagnetic (EM) waves [3]–[7]. On this basis, the performance of implantable antennas is a critical factor in enabling reliable wireless transmission, where the maximum gain and radiation pattern of implanted antennas are the key performance metrics [8], [9].

To quantitatively assess losses of implanted antennas caused by the biological tissues, an efficient approach is to use simplified body models that mimic realistic host bodies [10]–[17]. By employing these simplified models, such as spherical, cylindrical, or planar body models, the EM radiation process is analyzed to obtain design strategies or benchmarks for realistic implanted antennas. For instance, by means of the spherical body model, the fundamental bounds of the radiation efficiency are studied in [14], obtaining the optimal design of implantable antennas. With the aid of the planar body model, the optimal design for the wireless powering of medical implants containing a ferrite core is investigated in [15]. For large host bodies such as the human body, the planar body model can be used as an effective simplified model to analyze the effects of biological tissues with nearly-planar interfaces. Based on this model, this paper presents closed-form expressions to estimate the maximum gain and radiation pattern of implanted antennas within large host bodies. The proposed method considers multiple parameters, including the permittivity of the host body, implantation depth, operating frequency, and dimensions of the antenna encapsulation.

This paper is organized as follows. Section II describes the modeling of the implanted antenna and the derivation of closed-form expressions to estimate its radiation pattern and maximum gain. In Section III, we present the simulation and measurement results for multiple examples of implanted antennas to validate the proposed method as a fast and effective estimation method for link budget assessment before starting antenna design. Finally, conclusions are drawn in Section IV.

## II. SIMPLIFIED MODEL DERIVATION

### A. Modeling of an Implanted Antenna in a Large Host Body

To analyze the EM radiation from an antenna implanted in a large host body, we consider a canonical model in which an elementary dipole source is placed in a planar host body, as shown in Fig. 1. The preconditions for using this model to analyze implanted antennas are: the dimensions of the planar host body are electrically large, the host body has a nearly-flat body interface close to the implant, and the implantation depth of the source $d$ is significantly smaller than the distance from the source to other body interfaces.

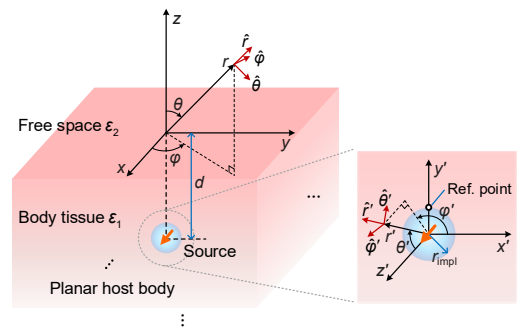

Fig. 1. A planar body model with an elementary dipole source implanted at a depth $d$.

The implanted antenna in the model is an electric dipole source surrounded by a small lossless sphere with a radius of $r_{impl}$, which roughly represents the dimension of the implant encapsulation. The large host body is modeled by a planar host body occupying the lower half-space. To simulate the lossy body tissue, the complex relative permittivity $\varepsilon_1$ derived from the four-region Cole–Cole model [18] is applied to the medium of the host body. The upper half-space is regarded as free space with $\varepsilon_2 = \varepsilon_0$. The source is implanted at a depth $d$ within the planar host body.

To characterize the radiation performance of the implanted antenna, its gain is of major concern. The latter comprises the losses due to the host body. With the knowledge of the far-field gain pattern, the link budget from the implanted antenna to an external access point can be directly estimated in the initial design stage of the link. To maximize the radiation gain, the orientation of the dipole source in the model is set to be parallel to the interface of the host body (i.e., an *x*-oriented dipole) such that the main radiation lobe of the dipole points towards the nearest body interface, as illustrated in Fig. 1. In the principal coordinate system for far-field gain analysis, the coordinate origin is placed above the dipole source at a distance $d$, i.e., at the interface of the host body in the $x$-$y$ plane.

*B. Closed-Form Expressions to Approximate Gain Patterns*

To estimate the far-field gain of the implanted antenna shown in Fig. 1, we need to analytically calculate the power density of EM waves, excited by the antenna, at the implant encapsulation and in the far-field region, respectively [19]. Note that the presence of a lossy host body causes the radiated power density to experience different levels of attenuation in different directions. As a result, decrease in radiation efficiency and deformation of the radiation pattern occur.

In dealing with EM fields surrounding the implant, an effective simplified method can be applied since most implant encapsulations have electrically small dimensions (with respect to the wavelength $\lambda$ in lossy tissue). According to the equivalence theorem [20], EM fields in the near-field region can be directly approximated by considering the equivalent antenna immersed in the body tissue but without encapsulation. In Fig. 1, a local coordinate system is introduced with the electric dipole source as the coordinate origin to facilitate the derivation of analytic expressions. A reference point is placed on the encapsulation closest to the body interface, which measures the maximum power density entering the lossy body tissue.

In the local spherical coordinate system, with applying spherical EM wave representation [20]–[23], the power density (i.e., the real part of the radial component of the Poynting vector $S_{r'}$) at $(r',\theta',\varphi')$ can be evaluated as

$$\mathrm{Re}\{S_{r'}\} = \mathrm{Re}\{E_{\theta'}H_{\varphi'}^*\}$$
$$= \sum_{n,m} \frac{C_{mn}}{r'^2}\left[\frac{d}{d\theta'}P_n^{|m|}(\cos\theta')\right]^2 \mathrm{Re}\left[j\zeta_1|k_1|^2 \hat{H}_n^{(2)}(k_1 r')\hat{H}_n^{(2)*}(k_1 r')\right], \quad (1)$$

where $\sum_{n,m} = \sum_{n=1}^{+\infty}\sum_{m=-n}^{n}$, $m$ and $n$ are the spherical mode indexes, $C_{mn}$ are constants related to the corresponding spherical modal coefficients, $P_n^m$ denotes the associated Legendre functions, $k$ is the complex wave number of the considered medium with a permittivity of $\varepsilon_r = \varepsilon_r' - j\varepsilon_r''$ and is calculated as $k = 2\pi f\sqrt{\varepsilon_r \varepsilon_0 \mu_0} = k' - jk''$, $\zeta$ is the intrinsic impedance of the considered medium, and $\hat{H}_n$ denotes the Schelkunoff spherical Hankel functions [20]. For the fundamental spherical mode $n = 1$ excited by the implant, corresponding to the mode with the lowest near-field loss [23], the power density at the reference point with radial coordinate $r_{\mathrm{impl}}$ (i.e., the maximum power density entering the body tissue) is simplified as

$$\mathrm{Re}\{S_{\mathrm{ref}}\}$$
$$= \left(\frac{Idl}{4\pi r_{\mathrm{impl}}}\right)^2 \mathrm{Re}\left\{\zeta_1\left[|k_1|^2 + 2k_1'' r_{\mathrm{impl}}^{-1} + \left(1-\frac{k_1^*}{k_1}\right)r_{\mathrm{impl}}^{-2} - jk_1^{-1}r_{\mathrm{impl}}^{-3}\right]\right\}e^{-2k_1'' r_{\mathrm{impl}}}, \quad (2)$$

where the dipole source is represented as an electric dipole of moment $Idl$, i.e., $C_{01}$ in (1) can be written as $(Idl/4\pi)^2$.

For the same electric dipole source implanted in a planar host body (i.e., the body model shown in Fig. 1), its radiation fields can be evaluated asymptotically by the Green's functions for layered medium [24]. At radio frequencies, the radiation fields of a horizontal electric dipole in a semi-infinite conducting medium have been investigated by Biggs [25]. The integral formulation of radiation fields is developed in terms of the vector and scalar potentials as

$$\mathbf{E} = -j\omega\mathbf{A} - \nabla V, \qquad \mathbf{H} = \frac{1}{\mu_0}\nabla\times\mathbf{A}. \quad (3)$$

Following the procedure similar to one developed by Sommerfeld (see e.g. [26], [27]), the vector potential due to the $x$-directed point source requires two components in order to satisfy the boundary conditions at the dielectric half-space interface. The first component $A_x$ is due to the excitation, and the second component $A_z$ is the effect of the interface of two dielectric media. It can be shown that these two components in the upper half-space (i.e. in the air) are equal

$$A_x = -j\mu_0 \frac{Idl}{4\pi}\int_{-\infty}^{+\infty}\frac{1}{k_{z1}+k_{z2}}e^{-jk_{z1}d}e^{-jk_{z2}z}H_0^{(2)}(k_\rho \rho)k_\rho dk_\rho, \quad (4)$$

$$A_z = -\mu_0 \frac{Idl}{4\pi}\cos\phi\frac{\partial}{\partial \rho}\int_{-\infty}^{+\infty}\frac{k_{z1}-k_{z2}}{k_1^2 k_{z2}+k_2^2 k_{z1}}e^{-jk_{z1}d}e^{-jk_{z2}z}H_0^{(2)}(k_\rho \rho)k_\rho dk_\rho, \quad (5)$$

where $k_{z,i}^2 = k_i^2 - k_\rho^2$ with $\mathrm{Re}(k_{z,i}) \geq 0$ and $\mathrm{Im}(k_{z,i}) \leq 0$ ensuring integration on the proper Riemann sheet, $\rho^2 = x^2 + y^2$, and $H_0^{(2)}$ stands for the zero-order Hankel function of the second kind.

We are interested only in the far-field components, thus

$$E_\theta = -j\omega(A_x \cos\theta\cos\phi - A_z \sin\theta), \quad E_\phi = j\omega A_x \sin\phi. \quad (6)$$

By calculating the Sommerfeld integrals in (4) and (5) using the steepest descent method, we obtain the E-field expression in the far-field region of the upper half-space

$$E_\theta \approx -j\frac{\zeta_2 k_2}{2\pi}Idl\frac{\cos\phi\cos\theta\sqrt{\varepsilon_1 - \sin^2\theta}}{\varepsilon_1 \cos\theta + \sqrt{\varepsilon_1 - \sin^2\theta}}e^{-jk_2\sqrt{\varepsilon_1-\sin^2\theta}\,d}\frac{e^{-jk_2 r}}{r}, \quad (7)$$

$$E_\phi \approx j\frac{\zeta_2 k_2}{2\pi}Idl\frac{\sin\phi\cos\theta}{\cos\theta + \sqrt{\varepsilon_1 - \sin^2\theta}}e^{-jk_2\sqrt{\varepsilon_1-\sin^2\theta}\,d}\frac{e^{-jk_2 r}}{r}. \quad (8)$$

The above expressions are similar to the asymptotic expressions derived in [25], i.e., they do not contain simplification of the square root terms.

On this basis, the power density at $(r,\theta,\varphi)$ becomes

$$\mathrm{Re}\{S_r\} = \mathrm{Re}\{E_\theta H_\varphi^* - E_\varphi H_\theta^*\} = \frac{1}{\zeta_2}\left(|E_\theta|^2 + |E_\varphi|^2\right). \quad (9)$$

So far, we have approximated the radiated power densities of EM waves from an antenna implanted in a planar host body, reaching the encapsulation and the far-field region in the upper free space, respectively. Therefore, by definition, the gain of an implanted antenna in dBi is

$$G_{\mathrm{dBi}} \approx G_{\mathrm{dBd}} + 1.76\ \mathrm{dB} = 10\log_{10}\frac{\mathrm{Re}\{S_r\}r^2}{\mathrm{Re}\{S_{\mathrm{ref}}\}r_{\mathrm{impl}}^2} + 1.76\ \mathrm{dB}, \quad (10)$$

where the radial distance $r$ can be canceled in the calculation, and 1.76 dB is the directivity of a short dipole.

For the model under analysis, the maximum gain of the implanted antenna appears in the $+z$-axis direction

$$G_{\mathrm{dBi,max}} = 10\log_{10}\frac{\mathrm{Re}\{S_{z,+z\text{-axis}}\}r^2}{\mathrm{Re}\{S_{\mathrm{ref}}\}r_{\mathrm{impl}}^2} + 1.76\ \mathrm{dB}, \quad (11)$$

where

$$\mathrm{Re}\{S_{z,+z\text{-axis}}\} = \mathrm{Re}\{E_{x,+z\text{-axis}}H_{y,+z\text{-axis}}^*\} = \frac{|E_{x,+z\text{-axis}}|^2}{\zeta_2}, \quad (12)$$

$$E_{x,+z\text{-axis}} \approx j60\frac{k_2^2}{k_1+k_2}Idl e^{-jk_1 d}\frac{e^{-jk_2 r}}{r}. \quad (13)$$

Finally, a simplified approximation of the maximum gain is

$$G_{\mathrm{dBi,max}} \approx 10\log_{10}\frac{(240\pi)^2 k_2^4 e^{-2k_1''(d-r_{\mathrm{impl}})}}{\zeta_2 \mathrm{Re}\left[\zeta_1\left(|k_1|^2 - jk_1^{-1}r_{\mathrm{impl}}^{-3}\right)\right]|k_1+k_2|^2} + 1.76\ \mathrm{dB}. \quad (14)$$

It is interesting to relate this result to an approach based on the spherical mode representation to determine the maximum power density (and therefore the maximum gain) given in [23] and [17]. The power density in the direction of maximum gain can be expressed by efficiency terms denoted as $\eta$:

$$\text{Re}\{S_{r,\max}\} = \frac{r_{\text{impl}}^2}{r^2}\text{Re}\{S_{\text{ref}}\} \cdot \eta_{\text{near field}} \cdot \eta_{\text{propagation}} \cdot \eta_{\text{reflection}} \cdot \eta_{\text{refraction}}, \quad (15)$$

where the expressions for the first three efficiencies are given in [23] and [17], and the expression for $\eta_{\text{refraction}}$ is given by

$$\eta_{\text{refraction}} \approx \lim_{\theta \to 0}\left(\frac{\sin\theta_1}{\sin\theta_2}\right)^2 = \left(\frac{k_2}{k_1'}\right)^2. \quad (16)$$

Here $\theta_1$ and $\theta_2$ are angles of incident and refracted EM wave at the body interface. The maximum gain derived in this way is equal

$$G_{\text{dBi,max}} \approx 10\log_{10}\frac{\text{Re}\{\zeta_1|k_1|^2\}e^{-2k_1'(d-r_{\text{impl}})}4\zeta_2 k_2^2}{\text{Re}\{\zeta_1\left(|k_1|^2 - jk_1^{-1}r_{\text{impl}}^{-3}\right)\}|\zeta_1+\zeta_2|^2\text{Re}\{1/\zeta_1\}k_1'^2} + 1.76\text{ dB}.$$

(17)

After rearranging the above expression, we get the final expression for the maximum gain:

$$G_{\text{dBi,max}} \approx 10\log_{10}\frac{4\zeta_2 k_2^4 e^{-2k_1'(d-r_{\text{impl}})}}{\text{Re}\{\zeta_1\left(|k_1|^2 - jk_1^{-1}r_{\text{impl}}^{-3}\right)\}|k_1+k_2|^2}\frac{|k_1|^2}{|k_1'|^2} + 1.76\text{ dB}.$$

(18)

It is noted that the difference between the expressions for the maximum gain obtained by two different approaches is the term $|k_1|^2/|k_1'|^2$. This term is a function of the loss tangent of the body tissue $\tan\delta_1 = \varepsilon_1''/\varepsilon_1'$, i.e., $|k_1|^2/|k_1'|^2 = 1/\cos^2(\delta_1/2)$. The value of this term is approximately equal to 1 for most biological tissues. Even for a high lossy body tissue with $\tan\delta_1 = 1$ (which is completely unrealistic), the difference between the two approaches is only 0.69 dB, while in the considered case of muscle tissue this difference is even less than 0.1 dB at 2.45 GHz (see appendix for a specific example comparison). Therefore, the two approximate approaches can obtain almost identical estimates of maximum gain.

Compared to the approximation by efficiency terms, the newly proposed approximation enables the assessment of the radiation pattern, which is more practical since the external units may not be fixed in the direction of maximum gain. Prior to the detailed design of implantable antennas, this approximation method has the advantage of quickly estimating the budget of wireless link from the implant to the external device at different directions in free space. It is possible to easily evaluate these closed-form expressions in a few seconds, without requiring the use of simulation tools.

III. VALIDATIONS AND APPLICATIONS

In this section, several examples of implanted antennas are analyzed to validate the proposed estimation method with full-wave simulations and measurements.

A. Simulation of a Dipole Antenna Implanted in a Planar Body Model

Regarding the implanted antenna model shown in Fig. 1, we use the full-wave simulation software FEKO to validate the results. FEKO is based on the Method of Moments (MoM) together with the surface equivalent principle needed to combine regions filled with different dielectric materials [28]. Specifically, in the simulation model, the lower half-space is filled with a medium having the complex permittivity of muscle [18], in which the implanted antenna is embedded. The implanted antenna is modeled as an elementary electric dipole located in a spherical air bubble of radius $r_{\text{impl}}$.

Multiple scenarios are analyzed in the simulation. The implantation depth of the antenna $d$, which is involved in the approximation of the far-field power density, is set to 3 or 5 cm, respectively. The encapsulation radius $r_{\text{impl}}$, which is involved in the approximation of the near-filed power density, is assigned to 2 or 5 mm, respectively. Two Industrial, Scientific, and Medical (ISM) bands, i.e., 434 MHz and 2.45 GHz, are selected as the operating frequencies. As shown in Fig. 2, the approximate radiation pattern calculated by the proposed method overlaps well with the simulated radiation pattern (deviations in all directions are within 0.2 dB) in each scenario. These small deviations are mainly due to the accurate antenna feed model used in the FEKO simulation.

The good agreement of these results verifies the effectiveness of the proposed method for implants within an ideal large host body, i.e., a planar host body occupying the lower half-space. Furthermore, it can be observed that an increase in implantation depth $d$ always reduces the gain value due to higher propagating field losses. At the same time, the encapsulation radius $r_{\text{impl}}$ has a more pronounced effect on the gain value at low frequencies, i.e., 434 MHz in Fig. 2. This is due to the fact that the electrical scale of the encapsulation is smaller at low frequencies, and the corresponding near-field losses of the implanted antenna become more dominant as the encapsulation radius $r_{\text{impl}}$ decreases. A detailed analysis of the near-field losses can be found in [17].

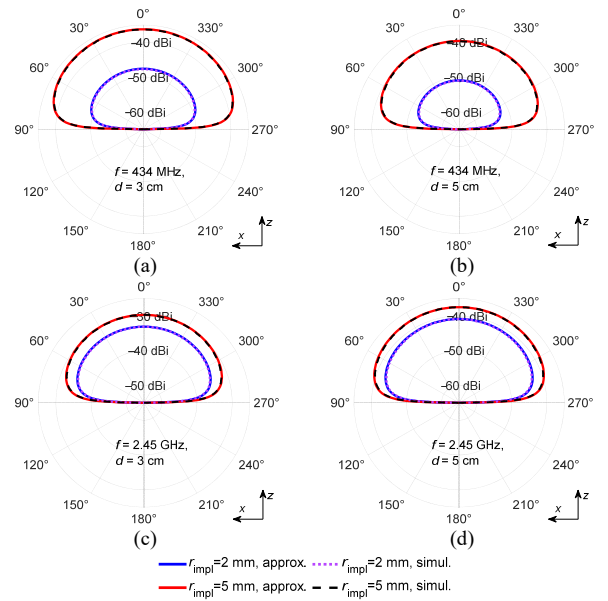

Fig. 2. Comparison between the approximate and simulated radiation patterns (in the E-plane) of the implanted dipole antenna in multiple scenarios: (a) $f$ = 434 MHz, $d$ = 3 cm; (b) $f$ = 434 MHz, $d$ = 5 cm; (c) $f$ = 2.45 GHz, $d$ = 3 cm; (d) $f$ = 2.45 GHz, $d$ = 5 cm. The encapsulation radius $r_{\text{impl}}$ in each subfigure is set to 2 and 5 mm, respectively.

B. Simulation of a planar inverted-F Antenna Implanted in a Planar Body Model

To demonstrate the practicality of the proposed method for a real implanted antenna, we study a 2.45-GHz planar inverted-F antenna (PIFA) implanted in a planar body model.

Specifically, this PIFA is encapsulated in a cuboidal medium with lossless dielectric ($\varepsilon_r$ = 4) representing a simplified model of a cardiac implant. The antenna is implanted in a planar body model made of muscle tissue with a depth of 3 cm. Fig. 3(a) shows the analyzed model and detailed geometry of the antenna.

We use CST Microwave Studio [29] to conduct full-wave simulation of the implanted antenna, and finally obtain the radiation pattern as shown by the solid black line in Fig. 3(b). The radiation pattern can also be estimated by the proposed approximation method. For this PIFA with a ground plane, its encapsulation radius $r_{impl}$ is approximated as 4 mm (i.e., with the thickness of the encapsulation). On this basis, the approximate radiation pattern is shown by the red dash line in Fig. 3(b). It can be observed that the approximation provides a robust assessment of radiation pattern for this real implanted antenna, which deviates from the simulated maximum gain by only –0.18 dB.

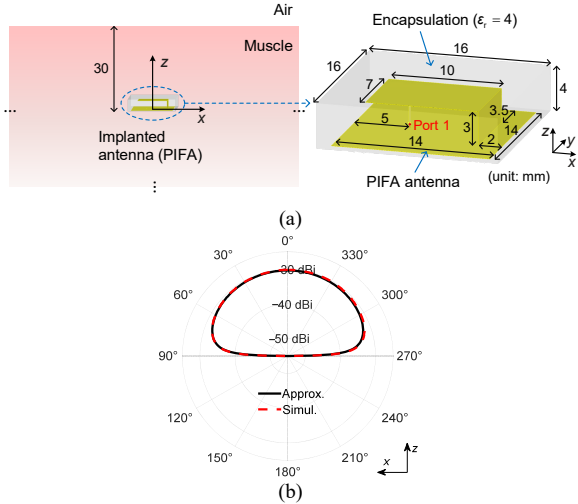

Fig. 3. (a)View of the PIFA antenna implanted in a planar body model along with detailed antenna geometry. (b) Comparison between the approximate radiation pattern (in the E-plane) and the corresponding simulated radiation pattern at 2.45 GHz.

### C. Simulation of a Capsule Antenna Implanted in the Abdomen of a Human Body Model

To validate the effectiveness of the proposed method in a realistic large host body, we investigate a 2.45-GHz capsule antenna implanted in the abdomen of a human body model.

The modeling and simulation of this implanted antenna example are carried out in full-wave simulation software CST Microwave Studio [29]. Specifically, in the simulation model shown in Fig. 4(a), a human body phantom (homogeneous pose library, CST Microwave Studio) is adopted and constructed of homogeneous body equivalent tissue with $\varepsilon_r = 29.01 - j6.77$ at 2.45 GHz [29]. A meander dipole antenna made of perfect electric conductor wire (radius 0.1 mm) is encapsulated in a capsule-shaped air bubble as the implanted antenna. Depending on the antenna dimensions shown in Fig. 4(a), the effective radius of the antenna $r_{impl}$ can be approximated as the circumradius of the encapsulation containing the antenna conductors [14], i.e., $r_{impl} \approx \sqrt{8.32^2/4 + 5.55^2/4} \approx 5$ mm. The antenna is implanted in the abdomen of the body model with the implantation depth $d = 3$ cm. The antenna is first oriented in parallel to the body interface, i.e., in the x-axis direction, corresponding to the maximum radiation gain. Then, to analyze the effect of rotation on the implanted antenna, the antenna is rotated by an angle of $\alpha$ in the horizontal plane, see Fig. 4(a).

For rotation angles $\alpha$ from 0° to 90° with a step of 30°, the simulated radiation patterns are demonstrated in Fig. 4(b), where the radiation pattern with the highest gain is, as expected, the scenario with $\alpha = 0°$ (i.e., an x-oriented antenna). Furthermore, the approximate radiation pattern via the proposed method is calculated, as shown by the red dashed curve in Fig. 4(b). The approximation results provide a good estimate of the maximum gain (deviation of only 0.13 dB) and the main lobe of the radiation pattern. This further validates the applicability of the proposed method to assess the radiation properties of realistic implants within large host bodies.

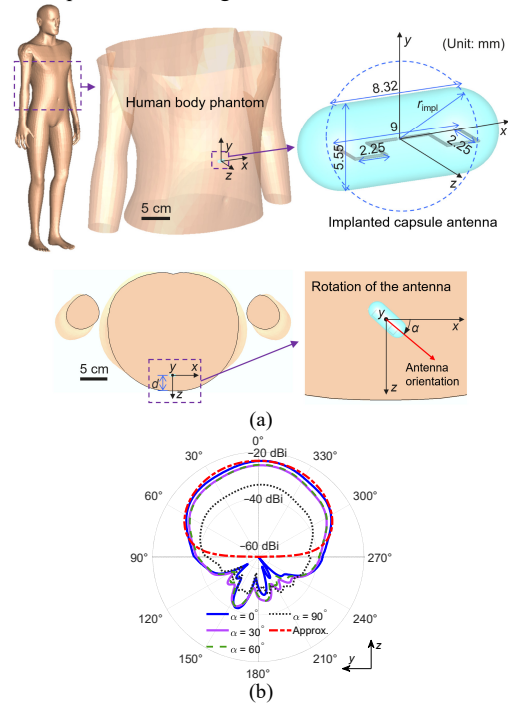

Fig. 4. (a) View of the capsule antenna implanted in the abdomen of a human body model, where the rotation angle of the implanted capsule antenna is denoted as $\alpha$. (b) Comparison between the approximate radiation pattern (in the H-plane) and the corresponding simulated radiation patterns (different rotation angles $\alpha$) of the implanted capsule antenna at 2.45 GHz.

### D. Measurement of a Ceramic-Encapsulated Antenna Implanted in a Cuboidal Liquid Body Phantom

The proposed approximation method is further validated by measuring a 2.45-GHz ceramic-encapsulated antenna in a cuboidal liquid body phantom.

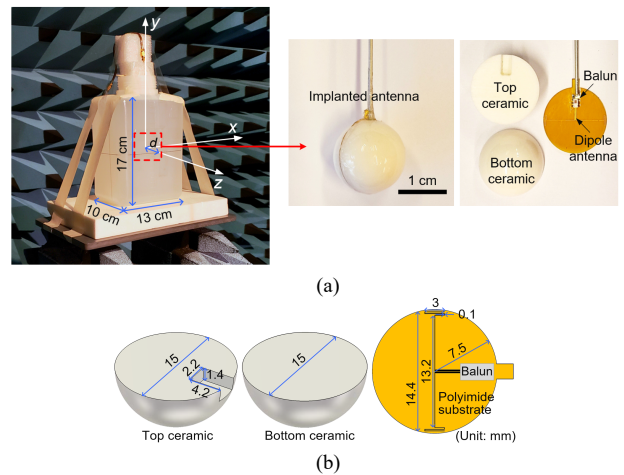

Fig. 5. (a) Photographs of the ceramic-encapsulated antenna implanted in a cuboidal liquid body phantom, where the antenna is immersed in the container and the dimensions are given. (b) Detailed structure and dimensions of the ceramic-encapsulated antenna.

In making a body phantom to mimic large host bodies, we take a cuboidal plastic (0.3 mm thick polyethylene, $\varepsilon_r$ = 2.3 – $j$0.0007) container filled with distilled water ($\varepsilon_r$ = 78 – $j$11.67 at 2.45 GHz) to represent the lossy body tissue, as can be seen in Fig. 5(a). The implantable antenna is a ceramic-encapsulated antenna consisting of a circular Flexible Printed Circuit (FPC) and two ceramic hemispheres as its encapsulation. On the FPC, a meander dipole antenna is designed and fabricated on a 0.1-mm thick flexible substrate (polyimide, $\varepsilon_r$ = 3.5 – $j$0.0025) with 18-μm thick copper metallization. A 2.4-GHz chip balun (2450BL15B050, Johanson Technology) is introduced in the transition from the unbalanced semi-rigid cable (EZ-34, EZ Form Cable) to the balanced feeding stripline of the dipole antenna. The FPC of the antenna is encapsulated by two ceramic hemispheres made of $Al_2O_3$ (alumina 99%, $\varepsilon_r$ = 9.8 – $j$0.001), which is low loss and biocompatible. The detailed parameters are given in Fig. 5(b).

An experimental setup for far-field measurements of the implanted antenna is built in an EM anechoic chamber, as shown in Fig. 5(a), where the implantation depth $d$ (i.e., the distance from the antenna to the closest container side wall) is 3.2 cm. The measured reflection coefficient |$S_{11}$| is shown in Fig. 6(a). At 2.45 GHz, the measured |$S_{11}$| is –17.18 dB, which is used to exclude mismatch loss in the gain measurement. A quad-ridge horn antenna (QH400, MVG Industries) is used as a reference antenna in the anechoic chamber, fixed at 2.1 m from the antenna under test. The measured radiation pattern is compared with the corresponding simulation pattern (via CST Microwave Studio) and the approximate pattern, as shown in Fig. 6(b). Note that the simulated results closely match the measured results, since both methods account for the cuboidal liquid body phantom. Although the proposed approximation method is derived from the analysis of an infinite-large planar body model, it can be found that the approximate results demonstrate an accurate estimation of the measured main lobe values in the radiation pattern. For instance, in the direction of 0°, the deviation of the antenna gain is only –0.19 dB between the measured and approximate results.

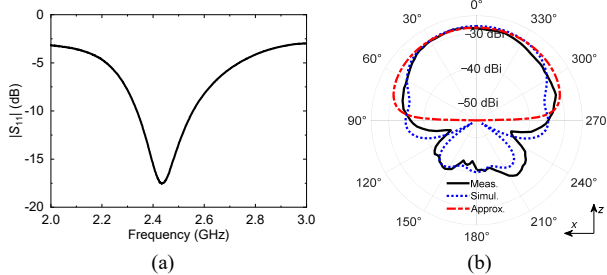

Fig. 6. (a) Measured |$S_{11}$| of the implanted ceramic-encapsulated antenna in a cuboidal liquid body phantom. (b) Comparison between the measured, simulated, and approximate patterns (in the E-plane) of the implanted antenna at 2.45 GHz.

## IV. CONCLUSIONS

In this paper, an estimation method has been developed that enables fast calculation of radiation patterns and maximum gain for antennas implanted in large host bodies. Simple closed-form approximate expressions are derived from the analysis of an electric dipole implanted within a planar host body, by means of spherical EM wave representation and Green's functions for layered medium. The key parameters of the implanted antenna, including the permittivity of the host body, implantation depth, encapsulation dimension, and operating frequency, are considered in the approximation.

To validate this newly proposed method, simulation, and measurement results from several implanted antenna examples are shown and analyzed, all of which match well with the approximate results. By enabling fast and accurate estimation of the maximum gain and radiation pattern of the implanted antenna, the proposed method can facilitate the development of more reliable and efficient wireless implantable devices.

## APPENDIX

To analyze the difference between the two approximate approaches (i.e., the proposed approximation and the approximation by efficiency terms) for maximum gain estimation, we consider the implanted antenna model shown in Fig. 1: an elementary electric dipole is located in a spherical air bubble of radius $r_{impl}$ = 5 mm, and this source is implanted at a depth of 3 cm within a planar body phantom. The operating frequency of the source is 2.45 GHz. In this analysis, we keep the real part of the body tissue permittivity $\varepsilon'_{r,1}$ as the value of muscle (i.e., $\varepsilon'_{r,1}$=52.73), but change the imaginary part of the body tissue permittivity $\varepsilon''_{r,1}$ from 0 to 60.

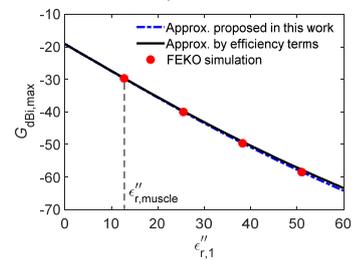

Fig. 7. Comparison of maximum gain calculated by two approximation approaches and FEKO simulation.

According to the results shown in Fig. 7, the difference between the maximum gain calculated by the two different approximate approaches is negligible even for high lossy body tissue, since it follows the property of $|k_1|^2/|k'_1|^2 = 1/\cos^2(\delta_1/2)$. Specifically, if the real and imaginary parts of the body tissue permittivity are equal (which is a very lossy medium and completely unrealistic), i.e., if $\varepsilon'_{r,1} = \varepsilon''_{r,1}$, the difference in results is still only 0.69 dB. For the imaginary part of the muscle permittivity (i.e., $\varepsilon''_{r,1} = 12.75$), the difference becomes only 0.06 dB. In addition, the same implanted antenna model is also simulated in FEKO, and the results for different values of imaginary part are given, which are also in good agreement with the approximate results.

## ACKNOWLEDGMENT

The authors are grateful to Prof. Juan R. Mosig for the useful discussions and guidance on Green's functions. The authors also thank Prof. Krzysztof A. Michalski for his valuable suggestions.

## REFERENCES

[1] D. Fitzpatrick, *Implantable Electronic Medical Devices*. New York, NY, USA: Elsevier, 2014.
[2] Q. H. Abbasi, M. U. Rehman, K. Qaraqe, and A. Alomainy, Eds., *Advances in Body-Centric Wireless Communication: Applications and State-of-the-art*. London, UK: IET, 2016.
[3] A. Kiourti and K. S. Nikita, "Implantable antennas: A tutorial on design, fabrication, and in vitro/in vivo testing," *IEEE Microw. Mag.*, vol. 15, no. 4, pp. 77–91, Jun. 2014.
[4] K. Agarwal, R. Jegadeesan, Y.-X. Guo, and N. V. Thakor, "Wireless power transfer strategies for implantable bioelectronics," *IEEE Rev. Biomed. Eng.*, vol. 10, pp. 136–161, Mar. 2017.

[5] N. A. Malik, P. Sant, T. Ajmal, and M. Ur-Rehman, "Implantable antennas for bio-medical applications," *IEEE J. Electromagn. RF Microw. Med. Biol.*, vol. 5, no. 1, pp. 84–96, Mar. 2021.

[6] I. V. Soares *et al.*, "Wireless Powering Efficiency of Deep-Body Implantable Devices," *IEEE Trans. Microw. Theory Tech.*, vol. 71, no. 6, pp. 2680-2692, June 2023.

[7] A. Basir *et al.*, "Implantable and Ingestible Antenna Systems: From imagination to realization [Bioelectromagnetics]," *IEEE Antennas Propag. Mag.*, vol. 65, no. 5, pp. 70-83, Oct. 2023

[8] M. Suzan Miah, A. N. Khan, C. Icheln, K. Haneda, and K.-I. Takizawa, "Antenna system design for improved wireless capsule endoscope links at 433 MHz," *IEEE Trans. Antennas Propag.*, vol. 67, no. 4, pp. 2687–2699, Apr. 2019.

[9] D. Nikolayev, A. K. Skrivervik, J. S. Ho, M. Zhadobov and R. Sauleau, "Reconfigurable Dual-Band Capsule-Conformal Antenna Array for In-Body Bioelectronics," *IEEE Trans. Antennas Propag.*, vol. 70, no. 5, pp. 3749-3761, May 2022.

[10] J. Kim and Y. Rahmat-Samii, "Implanted antennas inside a human body: Simulations, designs, and characterizations," *IEEE Trans. Microw. Theory Tech.*, vol. 52, no. 8, pp. 1934–1943, Aug. 2004.

[11] A. S. Y. Poon, S. O. Driscoll, and T. H. Meng, "Optimal frequency for wireless power transmission into dispersive tissue," *IEEE Trans. Antennas Propag.*, vol. 58, no. 5, pp. 1739–1750, 2010.

[12] M. Manteghi and A. A. Y. Ibraheem, "On the study of the near-fields of electric and magnetic small antennas in lossy media," *IEEE Trans. Antennas Propag.*, vol. 62, no. 12, pp. 6491–6495, Dec. 2014.

[13] D. P. Chrissoulidis and J.-M. Laheurte, "Radiation From an Encapsulated Hertz Dipole Implanted in a Human Torso Model," IEEE Trans. Antennas Propag., vol. 64, no. 12, pp. 4984–4992, Dec. 2016.

[14] D. Nikolayev, W. Joseph, M. Zhadobov, R. Sauleau, and L. Martens, "Optimal radiation of body-implanted capsules," *Phys. Rev. Lett.*, vol. 122, no. 10, p. 108101, Mar. 2019.

[15] D. K. Freeman and S. J. Byrnes, "Optimal Frequency for Wireless Power Transmission Into the Body: Efficiency Versus Received Power," *IEEE Trans. Antennas Propag.*, vol. 67, no. 6, pp. 4073-4083, June 2019.

[16] L. Berkelmann and D. Manteuffel, "Antenna Parameters for On-Body Communications With Wearable and Implantable Antennas," *IEEE Trans. Antennas Propag.*, vol. 69, no. 9, pp. 5377-5387, Sept. 2021.

[17] M. Gao, Z. Šipuš and A. K. Skrivervik, "Analytic Approximation of In-Body Path Loss for Implanted Antennas," *IEEE Open J. Antennas Propag.*, 2023.

[18] S. Gabriel, R.W. Lau, and C. Gabriel, "The dielectric properties of biological tissues: III. Parametric models for the dielectric spectrum of tissues," *Phys. Med. Biol.*, vol. 41, 2271, 1996.

[19] C. A. Balanis, *Antenna Theory: Analysis and Design*, 4th ed. Hoboken, NY, USA: Wiley, 2016.

[20] R. F. Harrington, *Time-Harmonic Electromagnetic Fields*. New York, NY, USA: McGraw-Hill, 1961.

[21] J. A. Stratton, *Electromagnetic Theory*. New York, NY, USA: McGraw-Hill, 1941.

[22] M. Bosiljevac, Z. Šipuš, and A. K. Skrivervik, "Propagation in finite lossy media: An application to WBAN," *IEEE Antenn. Wireless Propag. Lett.*, vol. 14, pp. 1546–1549, 2015.

[23] A. Skrivervik, M. Bosiljevac, Z. Šipuš, "Fundamental limits for implanted antennas: maximum power density reaching free space," *IEEE Trans. Antennas Propag.*, vol. 67, no. 8, pp. 4978–4988, Aug. 2019.

[24] A. Banos and J. P. Wesley, *The Horizontal Electric Dipole in a Conducting Half-Space*. Berkeley, CA, USA: Scripps Inst. Oceanography, Marine Physics Lab., Univ. California, Sep. 1953, ch. 4, Rep. 53–33.

[25] A. Biggs, "Radiation fields from a horizontal electric dipole in a semi-infinite conducting medium," *IRE Trans. Antennas Propag.*, vol. 10, no. 4, pp. 358–362, Jul. 1962.

[26] A. Sommerfeld, *Partial Differential Equations in Physics,* Academic Press, New York, 1972.

[27] J. R. Mosig and F. E. Gardiol, "A dynamical radiation model for microstrip structures," in *Advances in Electronics and Electron Physics*, vol. 59, P, Hawkes, Ed. New York: Acedemic Press, pp.139-237, 1982.

[28] *Altair FEKO*, Altair Engineering, Inc., www.hyperworks.com/feko

[29] *CST Microwave Studio*, CST Studio Suite, Dearborn, MI, USA, 2018.